# Investigating the Effect of Music and Lyrics on Spoken-Word Recognition

*Odette Scharenborg*[1,2] *and Martha Larson*[1,3,4]

[1] Centre for Language Studies, Radboud University Nijmegen, Netherlands
[2] Donders Institute for Brain, Cognition, & Behavior, Radboud University Nijmegen, Netherlands
[3] Institute for Computing and Information Sciences, Radboud University Nijmegen, Netherlands
[4] Intelligent Systems Department, Delft University of Technology, Netherlands

o.scharenborg@let.ru.nl, m.larson@let.ru.nl

## Abstract

Background music in social interaction settings can hinder conversation. Yet, little is known of how specific properties of music impact speech processing. This paper addresses this knowledge gap by investigating 1) whether the masking effect of background music with lyrics is larger than that of music without lyrics, and 2) whether the masking effect is larger for more complex music. To answer these questions, a word identification experiment was run in which Dutch participants listened to Dutch CVC words embedded in stretches of background music in two conditions, with and without lyrics, and at three SNRs. Three songs were used of different genres and complexities. Music stretches with and without lyrics were sampled from the same song in order to control for factors beyond the presence of lyrics. The results showed a clear negative impact of the presence of lyrics in background music on spoken-word recognition. This impact is independent of complexity. The results suggest that social spaces (e.g., restaurants, cafés and bars) should make careful choices of music to promote conversation, and open a path for future work.

## 1. Introduction

Music is an important part of the soundscape of social interaction settings. In bars, restaurants, and cafés, music serves to communicate information about the setting [1], thus creating an atmosphere. It also promotes conversational privacy [2]. However, the wrong soundscape choices may cause fatigue by increasing the effort necessary to carry on conversation [3], or even disrupt conversation entirely. This work contributes towards the goal of identifying the properties of background music that optimally allow conversations to continue unhindered in social settings. Despite the large body of work on the effect of the presence of background noise on speech processing (see for a review [4]), the influence of specific properties of music on speech processing is not well understood. Here, we focus on the impact of the presence of lyrics and of music complexity. We investigate the effect of music and lyrics on spoken-word recognition, which is known to be a central building block of speech perception (e.g., [5]).

Previous studies have established that music may interfere with speech processing [6],[7],[8],[9]. By masking acoustic information completely or partially, music and sung lyrics can make the speech signal less intelligible. This type of masking is called *energetic masking* [4],[10],[11],[12]. Energetic masking occurs due to the direct interaction of the background music and the speech signal in the same ear [10],[11]. The severity of the masking effect, and thus the reduction in intelligibility of the speech signal, is dependent on the number of "glimpses" still available to the listener [13]. "Glimpses" are time-frequency regions not masked by the background noise that can be used by the listener for speech recognition.

Further, *informational* masking [4],[10],[11],[12] can also occur. Informational masking is the remaining interference after the effect of energetic masking has been taken into account (e.g., [4],[11]). In our work, sources of informational masking can be the music itself, but also linguistic information in the form of lyrics. Given the ongoing neuroscience discussion on neural resources sharing between speech and music processing in the brain, cf., [14],[15], one could possibly expect both musical complexity and lyrics to interfere equally with speech perception. However, given the findings on the impact of speech background noise (e.g., [4]), it is also plausible that lyrics in music pose a unique problem for perception. Our work focuses on the questions: Is the masking effect of music with sung lyrics larger than that of music without lyrics; and what is the role of the music complexity?

To investigate these questions, a word identification experiment was set-up in which Dutch listeners listened to short, CVC Dutch words embedded in background music. Two listening conditions were created: the Lyrics condition (music with lyrics) and the Music-Only (music from the same song without lyrics). We expect a larger detrimental effect of the presence of lyrics in the background music on spoken-word recognition than when there are no lyrics present in the background music due to 1) an increase in energetic masking in the Lyrics condition compared to the Music-Only condition, and 2) a potential informational masking effect of the lyrics (where informational masking has a larger detrimental effect on intelligibility than energetic masking (at a similar SNR) [16]). Note, we do not exclude the possibility that other properties of music cause informational masking. Specifically, we also expect to observe effects related to music complexity.

Next, we cover related work, mentioning how we extend the current understanding of speech processing in background music. Then we explain our experimental set-up and results. Finally, we present an outlook on other specific properties of music promising for future study.

## 2. Related Work

Bars, restaurant and cafés are devoting increasing amounts of effort to designing their soundscape. The fact that music can be controlled [1], makes it a particularly important soundscape element. Work until now on music in restaurant settings has focused on its ability to mask other sounds e.g., [2]. This work proposes a music recommender system to support the choice of music that is an effective masker of speech noise.

Evidence that lyrics are a potential source of informational masking comes from studies that have investigated how lyrics in background music affect cognitive tasks. The impact of lyrical vs. non-lyrical music on foreign language vocabulary learning has been studied by [17]. This work found a short-term effect when the language of the sung lyrics was familiar to the learner. The impact of music on work attention was studied by [18]. This work recommends that music with lyrics should be avoided to avoid impact on worker efficiency.

We know of three studies that investigated the effect of background music on speech processing and included background music with lyrics [6],[7],[9]. However, none has investigated the role of lyrics, specifically. In contrast to preceding work, we isolate the effect of lyrics. Specifically, we aim to control for other factors in the music in our Music-Only and Lyrics conditions by using instrumental music and music containing lyrics taken from the same song.

## 3. Experimental set-up

### 3.1. Participants

Twenty native Dutch listeners (11 females; mean age = 24.9, SD = 5.1) from the Radboud University subject pool participated in the experiment. None of the participants reported a history of language, speech, or hearing problems. The participants were paid 5 Euros for their participation.

### 3.2. Materials

*3.2.1. Word stimuli*

The stimuli consisted of 150 Dutch CVC words spoken by a native speaker of Dutch, and were taken from an earlier study [19] investigating the role of word frequency and neighborhood density on native spoken-word recognition. The word frequency and neighborhood density of the 150 words, which were obtained from [20], were orthogonally varied (but not further investigated in this study).

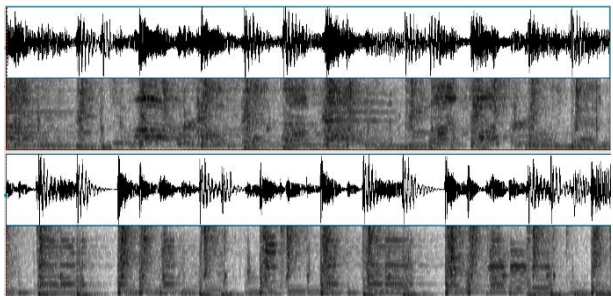

Figure 1. *Waveform and spectrogram of 4 seconds of the song "E go betta" (Song 1). Top panels with sung lyrics and bottom panels without lyrics.*

*3.2.2. Background music*

The CVC words were embedded in background music. Since our ultimate goal is to understand how music influences speech comprehension in bars and restaurants, we chose music from a specific restaurant in Amsterdam. A dedicated curator selects the music for the restaurant. The restaurant is popular and the curator ensures that the music songs both fit the atmosphere of the restaurant, and are fresh. By focusing on this restaurant, we could choose music that is varied in style, but not radically so. In this way, we could both ensure that we were experimenting with realistic restaurant music, and also minimize the impact of style differences between songs in our experiments. We chose the three songs listed in Table 1.

Table 1. *Bar/restaurant music used in the experiment*

|  | Name | Artist | Genre | Rhythm | bpm |
|---|---|---|---|---|---|
| Song 1 [21] | E go betta | Dele Sosimi | Afrobeat, Funk | complex | 110 |
| Song 2 [22] | Purple | Crustation | Down-tempo, Trip Hop | simple | 76 |
| Song 3 [23] | Stay away from music | Stephen Colebrook | Funk | simple | 118 |

In order to control as much as possible for the music instruments and the presence of beats in the songs between the Music-Only and Lyrics conditions, we chose songs that contained stretches with and without lyrics and where the instrumental music for these stretches was approximately the same. This was investigated by listening and visual inspection of the spectrograms of the songs. Of each song, two versions were created: one with and one without lyrics. The longest stretches with and without lyrics were selected from each song by carefully cutting the appropriate stretches on the positive-going zero-crossings using *Praat* [24]. We needed stretches of background music of approximately one minute in length. If no such stretch was present in the song, these were created by hand, by combining different stretches of the same song, while taking care that no abrupt changes in the music or lyrics would occur. This was checked both by listening and looking at the spectrograms. Figure 1 provides an example of a 4 seconds stretch for the Song 1 "E go betta". The top two panels show the condition with sung lyrics and the bottom two panels the condition without lyrics. As is clear, the overall structure of the music and beats is the same for the two conditions.

The resulting six background music files (3 songs, each in a Lyrics and a Music-Only condition) were then added to the stimuli at three different SNRs, i.e., SNR +15, +5, and 0 dB, using a custom-made *Praat* script. Each word stimulus was preceded by 200 ms of leading background music and followed by 200 ms of trailing background music. The stretch of background music was randomly selected from the background music files. A Hamming window was applied to the background music, with a fade in / fade out of 10 ms.

The SNRs were determined on the basis of a pilot study with 12 Dutch participants, none of whom participated in the current study. The SNRs were chosen such that for the easiest SNR, the background music is indeed perceived as being in the background, and at a level often found in coffee bars. The more difficult SNRs were chosen as to reflect a situation that is more to be expected in a pub or disco, as we were also interested in whether we could observe a point where the performance would 'break', i.e., would be severely impaired.

### 3.3. Procedure

Twelve experimental lists were created. Each list consisted of 150 items, with 50 items in each of the three SNR conditions. Half of the items in each SNR condition was assigned to the Lyrics condition (= 25 items per SNR condition) and the other half to the Music-Only condition. Finally, the three different songs were randomly assigned to one of the items. The order of the SNR and Lyrics/Music-Only blocks were randomized and counterbalanced across participants. Each participant was randomly assigned one list.

Participants were tested individually in a sound-treated booth. The stimuli were presented over closed headphones at a comfortable sound level. Participants listened to the 150 words and were asked to type in the word they thought they had heard. After pressing the return key, the next item was played.

## 4. Results

The top-left panel of Figure 2 shows the proportion of words correctly recognized for each of the SNR conditions for the two music backgrounds separately, averaged over the three songs. The dotted line shows the proportion correct for the music background without lyrics, the open-square line shows the proportion correct for the music background with lyrics. There was no performance difference for the easiest, 15 dB, listening condition, but Figure 2 shows a clear difference in recognition performance by the listeners between the two music conditions for the two more adverse listening conditions: fewer words were correctly recognized for the two worst listening conditions when the music contained lyrics compared to the condition where no lyrics were present.

Statistical analyses using generalized linear mixed-effect models (e.g., [25]), containing fixed and random effects, on the accuracy of the recognized words were carried out to investigate these observations. The dependent variable was whether the word stimulus was correctly identified ('1') or not ('0'). Fixed factors were SNR (3 levels: +15 dB (on the intercept), +5 and 0 dB; nominal variable, as this model (AIC=2513.9) significantly outperformed the model including SNR as a continuous variable (AIC=2526.0)), and crucially the absence (on the intercept) or presence of lyrics in the background noise. Stimulus, Subject, and Song were entered as random factors. Random by-Subject, by-Stimulus, and by-Song slopes for SNR were added, only the random by-Stimulus slope for SNR remained in the best-fitting model.

Table 2. *Fixed effect estimates for the best-fitting model for the overall analysis, n=3000.*

| Fixed effect | β | SE | p |
| --- | --- | --- | --- |
| Intercept | 2.958 | .448 | <.001 |
| SNR +5 | -.719 | .288 | .012 |
| SNR 0 | -1.121 | .312 | <.001 |
| Lyrics | -.050 | .235 | .83 |
| SNR +5 × Lyrics | -1.067 | .303 | <.001 |
| SNR 0 × Lyrics | -1.459 | .314 | <.001 |

Table 3. *Fixed effect estimates for the best-fitting model for Song 1, n=1000.*

| Fixed effect | β | SE | p |
| --- | --- | --- | --- |
| Intercept | 2.474 | .415 | <.001 |
| SNR +5 | -1.326 | .390 | <.001 |
| SNR 0 | -1.603 | .478 | <.001 |
| Lyrics | -.640 | .34 | .066 |
| SNR +5 × Lyrics | -.325 | .459 | .478 |
| SNR 0 × Lyrics | -1.375 | .523 | .009 |

Table 4. *Fixed effect estimates for the best-fitting model for Song 2, n=1000.*

| Fixed effect | β | SE | p |
| --- | --- | --- | --- |
| Intercept | 4.433 | .902 | <.001 |
| SNR +5 | -1.897 | .845 | .025 |
| SNR 0 | -1.910 | .879 | .030 |
| Lyrics | .640 | .561 | .254 |
| SNR +5 × Lyrics | -1.257 | .655 | .055 |
| SNR 0 × Lyrics | -2.488 | .681 | <.001 |

Table 5. *Fixed effect estimates for the best-fitting model for Song 3, n=1000.*

| Fixed effect | β | SE | p |
| --- | --- | --- | --- |
| Intercept | 2.491 | .380 | <.001 |
| SNR +5 | .832 | .556 | .135 |
| SNR 0 | -.450 | .529 | .395 |
| Lyrics | .381 | .399 | .340 |
| SNR +5 × Lyrics | -2.411 | .575 | <.001 |
| SNR 0 × Lyrics | -1.173 | .528 | .026 |

Table 2 shows the fixed effect estimates for the best-fitting model of the overall analysis. The statistical analysis confirmed the visual observations. Overall, at the two more adverse SNR levels, recognition accuracy was significantly worse than at the easier SNR level (SNR effects in Table 2). Regarding our crucial manipulation of the presence or absence of sung lyrics, for SNR +15 dB, there was no significant difference between the two music backgrounds, but as shown by the two interactions between SNR and Lyrics, significantly fewer words were recognized when the background music contained lyrics compared to when there were no lyrics at the two most difficult SNR conditions.

To investigate the influence of the songs, in particular that of rhythmic complexity, on spoken-word recognition, we carried out statistical analyses for the three songs separately. The top-right and bottom panels of Figure 2 show the proportion of words correctly recognized for each of the SNR conditions for the two music backgrounds for each of the songs separately.

Comparing the top right panel of Figure 2 (Song 1) and the bottom panels (Songs 2 and 3) shows that the word recognition performance for Song 1 was lower than that for the other two songs. Song 1 is thus an inherently better masker than Songs 2 and 3. We get back to this finding in the General Discussion. The key observation is that for all songs, the Music-Only condition outperforms the Lyrics condition.

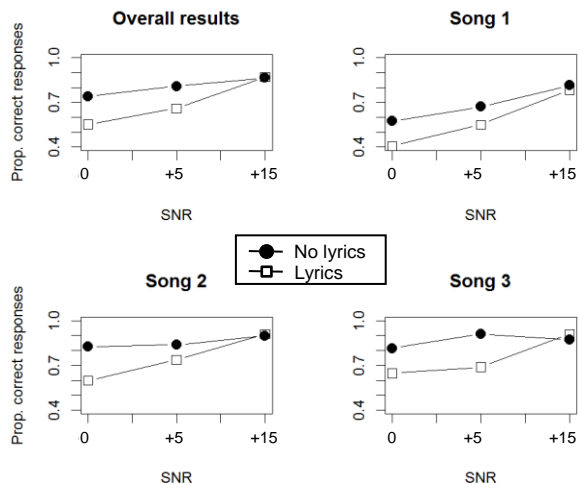

Figure 2. *Proportion of correct responses for the three SNR conditions for the two music backgrounds separately, averaged over the three songs (top left panel) and for each song separately (top right and bottom panels).*

Tables 2-4 show the fixed effect estimates for the best-fitting model of the per-song analyses. The biggest difference between the three songs is in the interactions between SNR and Lyrics. For all songs, at SNR 0 dB significantly fewer words were recognized when the background music contained lyrics compared to when there were no lyrics. For Song 3, this effect was also found for SNR +5 dB, and it was marginally present for Song 2 at SNR +5 dB.

## 5. General discussion

This work investigates the influence of the presence of lyrics and the complexity of the music in background music on spoken-word recognition. To that end, Dutch native listeners were tested on a CVC word-identification task in Dutch with background music, crucially with and without the presence of lyrics, at three different SNRs, and using three different songs with differences in music complexity. The key finding is that at the two worst SNR conditions, words in the Music-Only condition were significantly better recognized than words in the Lyrics condition. So, indeed, background music with sung lyrics has a larger masking effect on spoken-word recognition than background music without lyrics.

Additionally, differences were observed between the masking effects of the three songs that served as background music. Overall, listeners gave fewer correct answers to Song 1 than to the other two songs. The complex rhythm, involving swing timing, of Song 1 could be the source of a larger energetic masking effect compared to the music structures of Songs 2 and 3. Since the stimuli are 500-1000 ms in length, and the beats per minute (bpm) rates of the songs vary between 76–118 bpm, there are only 1–2 main beats per stimulus. However, for Song 1, many percussion notes are present between the main beats, as reflected by the spacing of the energy in Figure 1. This explanation would put the observation in line with findings from [8], which found a larger masking effect for faster tempos. Note that the tempo of Song 1 in bpm is slower than that of Song 3. However, the number of percussion notes heard by participants was effectively larger for Song 1. Future research will investigate the relationship between the proportion of 'glimpses' that are available to the listener [13] and the music complexity to get a grip on the amount of energetic masking caused by different music complexities.

The effect of the presence of lyrics was only found when the music was relatively loud in comparison to the target speech. The easiest listening condition did not show a difference in word recognition performance between the Music-Only and the Lyrics condition. Moreover, for Song 1 this effect was only found at the most difficult listening condition, while for the other two songs this difference was present for the two worst listening conditions (marginally so for Song 2), suggesting that the sung lyrics in Songs 2 and 3 are better maskers than the sung lyrics in Song 1. The relative energy of the singers' voices with respect to the other instruments in the song could play a role, and is an interesting perspective for future work. We note that the number of singers does not appear to impact speech perception. In contrast to Song 2 and 3, Song 1 has multiple vocalists singing in unison. An increase in background speakers results in an increase of the masking effect [16], so if the number of vocalists were to play a role, we would have expected a larger masking effect for Song 1 compared to Songs 2 and 3.

Finally, the results on the presence of lyrics appear to hint at a larger masking effect of a familiar language. The language in Song 1 appears to be West African Coastal English, which implies distinctive phonetics and possibly also an influence of tone. As such, the English of Song 2 and Song 3 is expected to be more familiar to the ears of the native Dutch language participants, who on a daily basis hear mostly English spoken in professional settings and Western entertainment. If the language of Song 2 and Song 3 is indeed more familiar, the results could point towards a larger masking effect of a known language. This finding would then be in line with results showing that listeners experience a larger masking effect from background babble when they understand the language of the speech in the background (e.g., [26],[27]). Note that due to the length of the stimuli, only word fragments and very rarely complete words are captured. For this reason, we would more readily expect an effect due to the phonetics of the lyrics language than an effect of familiarity with the language.

## 6. Conclusions and outlook

To our knowledge, this is the first study investigating the effect of background music with and without lyrics on spoken-word recognition. Our experimental results extend existing knowledge on the effect of different masker types on spoken-word recognition. Importantly, they also provide a baseline for the impact of background music on conversation in social settings. On one hand, isolated words are more difficult to recognize than words in context, so the adverse effects we observed could be expected to be worse than in more natural conversational settings. On the other hand, words are easier to recognize in a carefully controlled lab situation where the stimuli are played over headphones in a sound-proof booth compared to a more natural listening setting where listeners are typically at a (small) distance from one another.

The process of designing the experiment to isolate the impact of lyrics led to an interesting list of other factors that potentially influence how music affects conversations in social settings. Above, we already mentioned listener familiarity with the language of the lyrics, and the relative sound power of the singers with respect to the instruments as important. Additionally, there are factors that are related to the ability of listeners to separate streams of sounds. In our study, Song 1 had swing timing, and could be perceived as less predictable to listeners than Song 2 or 3, with straight timing. The ability to separate streams has been related to speech comprehension [28]. To understand how listeners' ability to anticipate the rhythm impacts word recognition, we can move, in the future, to longer samples with more than 1-2 main beats. Further, the age of the listener is also expected to play a role, cf. [6],[7], as well as the musical background of the listener, familiarity with the genre, and familiarity with the specific song cf. [6],[7],[29].

## 7. Acknowledgements

Odette Scharenborg was sponsored by a Vidi-grant from NWO (grant number: 276-89-003). Martha Larson was supported in part by EU FP7 project no. 610594 (CrowdRec). The authors would like to thank the student assistants of the lab of O.S. for help in setting up and running the experiments. We would also like to thank Kollekt.fm and Coffee and Coconuts for helping us understand music relevant for social interaction environments, and providing us with sample songs.

# 8. References

[1] P.M. Lindborg, "A taxonomy of sound sources in restaurants", *Applied Acoustics*, vol. 110, pp. 297-310, 2016.

[2] T. Kato, M. Oka and H. Mori, "Music recommendation system to be effectively difficult to hear the speech noise", *The SICE Annual Conference*, Japan, pp. 2353-2359, 2013.

[3] J.H. Rindel, "Verbal communication and noise in eating establishments", *Applied Acoustics*, vol. 71, pp. 1156-1161, 2010.

[4] M.L.G. Garcia Lecumberri, M. Cooke, and A. Cutler, "Non-native speech perception in adverse conditions: A review", *Speech Communication*, vol. 52, pp.864-886, 2010.

[5] J.M. McQueen, "Speech perception", In K. Lamberts & R. Goldstone (Eds.), *The handbook of cognition* (pp. 255-275). London: Sage Publications, 2004.

[6] F. Russo and M.K. Pichora-Fuller, "Tune in or tune out: Age-related differences in listening to speech in music", *Ear Hear.*, vol. 29, pp. 746-760, 2008.

[7] D. Başkent, S. van Engelshoven, and J.J. Galvin, "Susceptibility to interference by music and speech maskers in middle-aged adults", *The Journal of the Acoustical Society of America*, vol. 135, EL147, 2014.

[8] S. Ekstrom and E. Borg, "Hearing speech in music", *Noise Health* vol. 13, pp. 277-285, 2011.

[9] K. Gfeller, C. Turner, J. Oleson, S. Kliethermes, and V. Driscoll, "Accuracy of cochlear implant recipients on speech reception in background music", *Ann. Otol. Rhinol. Laryngol.*, vol. 121, pp. 782-791, 2012.

[10] M. Cooke, M.L. Garcia-Lecumberri, and J. Barker, "The foreign language cocktail party problem: Energetic and informational masking effects in non-native speech perception", *Journal of the Acoustical Society of America.*, vol. 123, no. 1, pp. 414-27, 2008.

[11] S. Mattys, J. Brooks, and M. Cooke, "Recognizing speech under a processing load: Dissociating energetic from informational factors", *Cogn. Psych.,* vol. 59, pp. 203-243, 2009.

[12] B.G. Shinn-Cunningham, "Object-based auditory and visual attention", *Trends in Cognitive Sciences,* vol. 12, pp. 182-186, 2008.

[13] M. Cooke, "A glimpsing model of speech perception in noise", *Journal of the Acoustical Society of America*, vol. 119, pp. 1562-1573, 2006.

[14] A.D. Patel, "Language, music, syntax and the brain", *Nature Neuroscience*, vol. 6, pp. 674-681, 2003.

[15] R. Kunert and L.R. Slev,. "A commentary on: "Neural overlap in processing music and speech", *Front. Hum. Neurosci.*, vol. 9, pp. 330, 2015.

[16] S. Simpson and M. Cooke, "Consonant identification in N-talker babble is a nonmonotonic function of N", *J. Acoust. Soc. Am.*, vol. 118, pp. 2775–2778, 2005.

[17] A.M.B. de Groot and H.E. Smedinga, "Let the music play! A short-term but no long-term detrimental effect of vocal background music with familiar language lyrics on foreign language vocabulary learning", *Studies in Second Language Acquisition*, vol. 36, no. 4, pp. 681-707, 2014.

[18] Y.-N. Shih, R.-H. Huang, and H.-Y. Chiang. "Background music: Effects on attention performance". *Work 42*, pp. 573-578, 2012.

[19] F. Hintz and O. Scharenborg, "Effects of frequency and neighborhood density on spoken-word recognition in noise: Evidence from spoken-word identification in Dutch", *Architectures and Mechanisms for Language Processing (AMLaP)*, Bilbao, Spain, 2016.

[20] V. Marian, J. Bartolotti, S. Chabal, and A. Shook, "CLEARPOND: Cross-Linguistic easy-access resource for phonological and orthographic neighborhood densities", *PLoS ONE*, Vol. 7, no. 8, 2012.

[21] https://www.youtube.com/watch?v=UMZ6TDrsOO0 (accessed March 2017)

[22] https://www.youtube.com/watch?v=otzbG7iZ0hc (accessed March 2017)

[23] https://soundcloud.com/djfryer/ath032b-stephen-colebrook-stay-away-from-music (accessed March 2017)

[24] P. Boersma, and D. Weenink, D. "Praat: doing phonetics by computer [Computer program]", 2013. Retrieved from http://www.praat.org/

[25] R.H. Baayen, D.J. Davidson, and D.M. Bates, "Mixed-effects modeling with crossed random effects for subjects and items", *Journal of Memory and Language*, vol. 59, pp. 390-412, 2008.

[26] M.L. Garcia Lecumberri, and M. Cooke, "Effect of masker type on native and non-native consonant perception in noise", *Journal of the Acoustical Society of America*, vol. 119, no. 4, pp. 2445-2454, 2006.

[27] K.J. Van Engen, "Similarity and familiarity: Second language sentence recognition in first- and second-language multi-talker babble", *Speech Communication*, vol. 52, no. 11-12, pp. 943–953, 2010.

[28] L.-F. Shi and Y. Law, "Masking effects of speech and music: Does the masker's hierarchical structure matter?" *International Journal of Audiology*, 49, pages 296-308, 2010.

[29] N. Perham and H. Currie, "Does listening to preferred music improve reading comprehension performance? *Applied Cognitive Psychology, Appl. Cognit. Psychol.*," vol. 28, 279-284, 2014.